\documentclass[prc,amsmath,twocolumn,showpacs,superscriptaddress]{revtex4}
\usepackage[dvips]{graphicx}
\usepackage{dcolumn}

\begin{document}

\title{Core excitation in the elastic scattering and breakup of $^{11}$Be on protons}

\author{N.~C.~Summers}
 \affiliation{Department of Physics and Astronomy, 
  University of Tennessee, Knoxville, Tennessee 37996}
 \affiliation{Department of Physics and Astronomy, 
  Rutgers University, Piscataway, New Jersey 08854}
 \affiliation{National Superconducting Cyclotron Laboratory,
Michigan State University, East Lansing, Michigan 48824}
\author{F.~M.~Nunes}
 \affiliation{National Superconducting Cyclotron Laboratory,
Michigan State University, East Lansing, Michigan 48824}
 \affiliation{Department of Physics and Astronomy,
Michigan State University, East Lansing, Michigan 48824}
\date{\today}

\begin{abstract}
The elastic scattering and breakup of $^{11}$Be from a proton target 
at intermediate energies is studied. We explore the
role of core excitation in the reaction mechanism. 
Comparison with the data suggests that
there is still missing physics in the description.
\end{abstract}


\maketitle

\section{Introduction}

Nuclear reactions offer the most diverse methods to study nuclei
at the limit of stability.  Understanding reaction mechanisms in
nuclear processes involving nuclei near the driplines is of great
importance, particularly at this time, when there is such a high
demand for accuracy on the structure information to be extracted
from the data. Reaction and structure models are undoubtedly
entangled, therefore improving reaction models often implies
incorporating more detailed structure models in the description
\cite{review}.

It is generally accepted that, in reactions with loosely bound
nuclei, the coupling to the continuum needs to be considered.
Continuum effects are very much enhanced in breakup but can also
have imprints on other reaction channels, for example elastic and
inelastic scattering. One framework that explicitly includes
continuum effects is the Continuum Discretized Coupled Channel
(CDCC) method \cite{cdcc}. A large amount of work has been devoted
to the analysis of experiments within this framework
\cite{nunes99,tost02,moro03,matsu04,ogata06} and in general
results are very good.

The eXtended Continuum Discretized Coupled Channel (XCDCC) method
\cite{xcdcc1,xcdcc2} was recently developed. It brings together a
coupled channel description of the projectile with a coupled
channel model of the reaction, enabling the description of
interference between the multichannel components of the projectile
as well as dynamical excitation of the core within the projectile,
during the reaction \cite{xcdcc2}. The model has been applied to
the breakup of $^{11}$Be$\rightarrow ^{10}$Be$+n$ and 
$^{17}$C$\rightarrow ^{16}$C$+n$ on $^9$Be at $\approx 60$
MeV/nucleon \cite{xcdcc1,xcdcc2}. The projectiles are described within a
two-body $core+n$ multi-channel model, where the core can be in
the ground state but also in an excited state. This model produces
breakup cross sections to specific final states of the core, given
a coupled channel Hamiltonian for the projectile. Results
presented in Ref.~\cite{xcdcc1,xcdcc2} show that core excitation
effects in the total cross section to the ground state of the core
are small, but become very large when considering the total
population of the core's excited state. Other differences can be
seen in angular and energy distributions but at present no such
data is available. The effect of core excitation needs to be
studied in other regimes, for very light and very heavy targets,
as well as a variety of energies.

In this work we concentrate on the proton target. In this case the
process is nuclear driven, and recoil effects are very important.
Several reactions of $^{11}$Be on protons have been measured in a
number of facilities, namely elastic scattering at 50 MeV/nucleon \cite{cortina}, 
quasi-elastic and breakup at 64 MeV/nucleon \cite{shrivastava} as well as transfer at 35 MeV/nucleon \cite{winfield}.

$^{11}$Be proton elastic scattering  at 49.3 MeV/nucleon was
performed in GANIL \cite{cortina}, at the same time as the elastic
scattering of the core $^{10}$Be at 59.4 MeV/nucleon.
Even though the outcoming $^{11}$Be measurements correspond to quasi-elastic, 
these are essentially elastic as the contribution from the first excited state is negligible.
Standard optical potentials (either density folding as in JLM \cite{jlm}
or global optical potentials coming from elastic fits as in CH89 \cite{ch89}) 
could reproduce the $^{10}$Be elastic reasonably well, requiring small 
renormalizations of the real and imaginary  
parts of the interaction ($\lambda_V=0.9$ and $\lambda_W=1.1$ for CH89) \cite{cortina} .
Larger renormalizations were required in order to reproduce the
distribution of $^{11}$Be ($\lambda_V=0.7$ or $\lambda_W=1.3$ for CH89) \cite{cortina}.

It is clear from Ref.~\cite{cortina} that the global optical potential
overestimates the elastic cross section for $^{11}$Be.
In Ref.~\cite{johnson97} the elastic scattering of $^{11}$Be on $^{12}$C 
was successfully described using a $^{10}$Be+n two-body model, incorporating breakup effects. 
As the $^{10}$Be-target interaction was fixed by the $^{10}$Be elastic scattering data, 
the large modification in the $^{11}$Be$+^{12}$C elastic data was described, 
without renormalization, purely through breakup effects.
Due to the loosely bound nature of the last neutron in $^{11}$Be this loss of flux 
from the elastic channel can be attributed to breakup into $^{10}$Be+n.
It is thus possible that the large renormalizations for $^{11}$Be scattering on 
protons \cite{cortina} are also due to breakup effects.

In Ref.~\cite{shrivastava}, elastic data is only described after large renormalizations of 
both the real and imaginary part of the $^{10}$Be+p interaction 
($\lambda_V=0.75$ and $\lambda_W=1.8$), much larger than those used in Ref.~\cite{cortina}.
These same renormalizations can no longer describe 
the $^{10}$Be+p elastic data from Ref.~\cite{cortina}, and are inconsistent with few-body reaction theory.
We will re-examine the elastic scattering of $^{10/11}$Be+p to see if one can 
consistently describe both sets of data using the same interaction  for $^{10}$Be+p,
by including continuum and core excitation.

In addition to elastic and inelastic measurements, breakup data from NSCL exist at 63.7 MeV/nucleon \cite{shrivastava}.
This breakup data is integrated into two wide energy bins due to statistics.
The lower energy bin covers the 1.78 MeV resonance and a reasonable angular distribution 
is obtained, which underpredicts the cross section \cite{shrivastava}.
The higher energy bin covers resonances that are thought to be built on excited core states.
The calculations presented in Ref.~\cite{shrivastava} failed to reproduce the shape of this higher energy bin, and the authors suggest that the source of the disagreement may be due to an active core during the reaction.
Now that it is possible to include core excitation in the reaction mechanism 
\cite{xcdcc2} we will re-examine the breakup data using a consistent $^{10}$Be+p interaction, 
and including systematically the  coupling to the $2^+$ state in $^{10}$Be.

Transfer reactions have also been performed with the $^{11}$Be
beam, at 35.3 MeV/u in GANIL \cite{winfield} with the aim of
extracting spectroscopic factors for the ground state. While the
reaction mechanisms proved to be more complicated than the 1-step
DWBA theory, results for (p,d) show evidence for a significant
core excited component. The inverse reaction,
$^{10}$Be(d,p)$^{11}$Be, has also been studied in GANIL
\cite{delaunay}, the main interest being the resonance structure
of $^{11}$Be. This illustrates how transfer is being used beyond
the standard application of spectroscopy of bound states,
underlining the need to better understand the transfer mechanism
and its coupling to the continuum.

All these different data offer a good testing ground for theory. A
comprehensive theoretical study \cite{keeley} focusing on
$^{10}$Be(d,p) show inconsistencies of the extracted spectroscopic
factors for data at different energies. Optical potential
uncertainties and core excitation effects could be at the heart of
the problem. 

In this work we perform calculations including elastic, inelastic, and breakup channels of $^{11}$Be on protons at intermediate energies.
We explore explicitly the effect of the inclusion of core excitation in the reaction mechanism. Comparison to elastic and breakup data will be presented here.
The analysis of the inelastic channel is presented in \cite{pain} and we leave a detailed study  of the transfer channel for a future publication. 
In section \ref{details} we provide the details of the calculations.
In section \ref{results} we present the results: first for the elastic channel (\ref{results:el}), then for the breakup (\ref{results:bu}).
Finally, in section \ref{conclusion}, we draw our conclusions and provide an outlook into the future. 

\section{\label{details} Details of the calculations}

The calculations for breakup of loosely bound systems on protons have a rather
different convergence requirement as compared to the breakup on heavier systems.
The model space needs to span large excitation energies, while the radial dependence
can be reduced significantly.  

For the CDCC calculations at 40 MeV/nucleon, the continuum was discretized upto 35 MeV, with 10 bins upto 10 MeV for $s$-, $p$-, and $d$-waves, and 8 bins from 10--35 MeV. We include 12 bins from 0--35 MeV for all other partial waves up to $l_{\mathrm{max}}=4$. The same binning scheme was used for the XCDCC calculations, except that the higher bin density upto 10 MeV was only used for channels with outgoing ground state core components.
Partial waves up to $l_{\mathrm{max}}=4$ were used for the coupled channels projectile states.

For the 60 MeV/nucleon CDCC calculations, a slightly different binning scheme was adopted to match the experimental energy bin integrations.
From 0--0.5 MeV, 2 bins were used; over the observed energy bins from 0.5--3.0 and 3.0--5.5 MeV, 3 bins were used in each case; and from 5.5--30 MeV, 6 bins.
For the XCDCC calculations where the outgoing channel had excited core states, only 1 bin was used from 0--0.5 MeV, 1 bin for each observed energy range, and 5 bins above.

The radial integrals for the bins were calculated upto 40 fm in steps of 0.1 fm.
The radial equations in the CDCC method were calculated for 30 partial waves with the lower radial cutoff for the integrals set to 4 fm inside the point Coulomb radius, and matched to the asymptotic Coulomb functions at 150 fm.

The $^{11}$Be bound state potential parameters are taken from Ref.~\cite{be12nunes}, using the Be12-pure interaction for the CDCC calculations and the Be12-b for the XCDCC calculations.
The $^{10}$Be-proton interaction is fitted to the proton elastic data available at the two energies.
A good fit could be obtained from a renormalized CH89 interaction \cite{ch89}. 
The parameters are given in Table~\ref{TABLE:pot}.
For the cases including $^{10}$Be excitation, 
the OM potentials were deformed with the same $\beta_2$ deformations as used in the $^{11}$Be bound state.
The coupling matrix elements to the excited state in $^{10}$Be assume a rotational model with the deformation fitted  to the experimental $B$(E2) strength \cite{be11nunes}.
The deformation length is in good agreement with that obtained from inelastic scattering ot the $2^+$ state in $^{10}$Be \cite{iwasaki00},
and the optical potential used here reproduces the angular distribution of the inelastic scattering well.

\begin{table}\begin{tabular}{|c|cccccc|}\hline
 energy & $V$   & $R_V$ & $a_V$ & $W$   & $R_W$ & $a_W$ \\ \hline
     40 & 60.84 & 1.000 & 0.7   & 23.16 & 0.600 & 0.6   \\
     60 & 31.64 & 1.145 & 0.69  &  8.78 & 1.134 & 0.69  \\ \hline
\end{tabular}
\caption{\label{TABLE:pot}$^{10}$Be-proton Woods-Saxon potential parameters. All energies are in MeV and lengths in fm.}
\end{table}

\section{\label{results} Results}

\subsection{\label{results:el} Elastic channel}

\begin{figure}
\includegraphics[height=5cm]{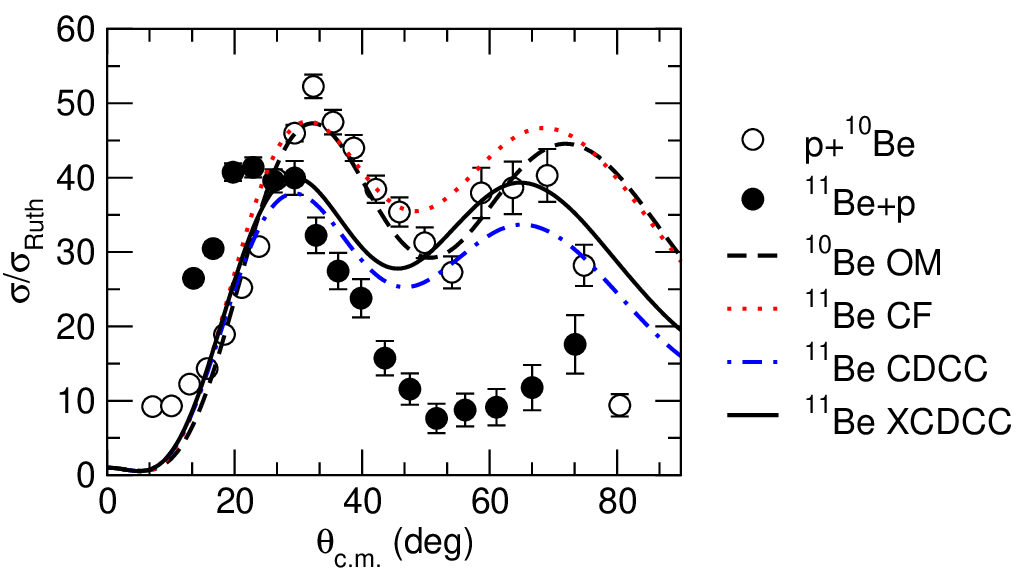}
\caption{\label{FIG:be-el-40} (Color online)
$^{10/11}$Be elastic scattering on a proton target at $\sim$40 MeV/nucleon, using an optical model fit to the $^{10}$Be elastic and various $^{11}$Be reaction models for the $^{11}$Be data.
The experimental data are from GANIL \cite{GANIL-unpub}.}
\end{figure}
\begin{figure}
\vspace{0.5cm}
\includegraphics[height=5cm]{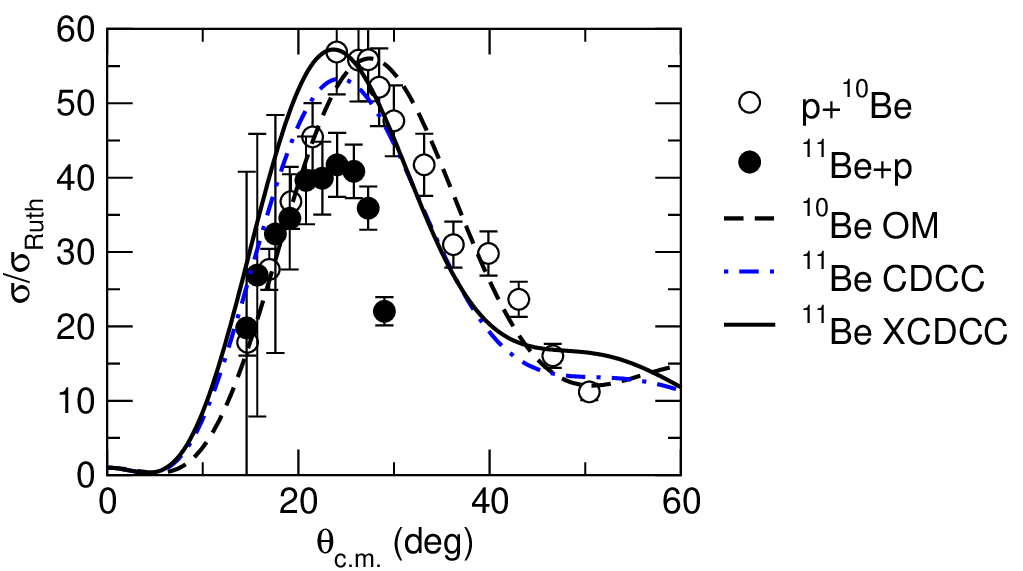}
\caption{\label{FIG:be-el-60} (Color online)
$^{10/11}$Be elastic scattering on a proton target at $\sim$60 MeV/nucleon, using an optical model fit to the $^{10}$Be elastic and various $^{11}$Be reaction models for the $^{11}$Be data. The experimental data are from Ref.~\cite{cortina} ($^{10}$Be) and Ref.~\cite{shrivastava} ($^{11}$Be).}
\end{figure}

The elastic scattering is the first test on the reaction model.
In Fig.~\ref{FIG:be-el-40} we show the $^{10}$Be and $^{11}$Be
elastic data and theoretical calculations at $\sim$40 MeV/nucleon. 
The optical model for the $^{10}$Be (dashed/black line) is fitted to the $^{10}$Be elastic data (open circles).
The cluster folding model (dotted/red line) folds the $^{10}$Be+p and n-p interactions over the $^{11}$Be ground state wave function to produce the $^{11}$Be+p potential.
Also shown in Fig.~\ref{FIG:be-el-40} is the effect of the $^{11}$Be continuum within CDCC (dot-dashed/blue line)  and core excitation within XCDCC (solid line).
Even though there is significant improvement over the simple optical model when including breakup, results still over-estimate the $^{11}$Be elastic cross section at larger angles, and no improvement is found by including excited core contributions.

Calculations were repeated for a higher energy, around 60 MeV/nucleon, where both $^{10}$Be and $^{11}$Be elastic data exist.
Once again, when the $^{10}$Be data is fitted with an optical model,
and the $^{11}$Be elastic is described within the CDCC approach, the cross section is over-estimated (see Fig.~\ref{FIG:be-el-60}). 
Note that the data at this energy does not span a large angular range, but it is evident that the pattern of over-predicting the $^{11}$Be cross section remains.

Other reaction calculations have been performed in an effort to describe this data \cite{crespo}, which consisted of a transfer to the continuum approach in which the breakup continuum was described using the deuteron basis.
This also failed to describe the data when the $^{10}$Be potential was fixed to the elastic data.
The same pattern of overpredicting the $^{11}$Be elastic was also seen at a lower energy of $\sim$40 MeV/nucleon \cite{crespo}.

As pointed out earlier, in \cite{cortina,shrivastava} large renormalization factors were needed to reproduce the elastic cross section. By including more relevant reaction channels, one might account for a part of the renormalization required, corresponding to flux that is being removed from the elastic channel. 
This suggests that there are still channels coupled to the elastic that have not been considered.
Preliminary calculations including the deuteron transfer channel along with the breakup in the $^{11}$Be basis show improvement at small angles, but the disagreement still remains at large angles.
Due to large non-orthonality corrections, CDCC calculations including the deuteron transfer
coupling turn out to be numerically challenging. They will be discussed in a later publication.

\subsection{\label{results:bu} Breakup: comparison with data at $\sim$60 MeV/nucleon}

Breakup data was also obtained at 63.7 MeV/nucleon \cite{shrivastava}, summed into two energy bins. The first covers the energy range 0.5--3.0
MeV, which spans the 1.78 MeV resonance, predominantly a $d$-wave
neutron coupled to the ground state of the core. The second energy
bin is over the energy range 3.0--5.5 MeV, which spans a resonance
at 3.89 MeV, thought to be predominantly an $s$-wave neutron coupled
to a $^{10}$Be($2^+$) core \cite{shrivastava}. In
Ref.~\cite{shrivastava}, CDCC results were presented 
which underestimated the cross section for the lower energy bin, but did not
reproduce the higher energy bin. It was suggested that since the
higher energy bin spanned a resonance with a possible
excited core component, the disagreement could be due to the
spectator core approximation in the standard CDCC theory. Since
XCDCC can handle excited core components, this data is
re-examined.

\begin{figure}
\includegraphics[height=5cm]{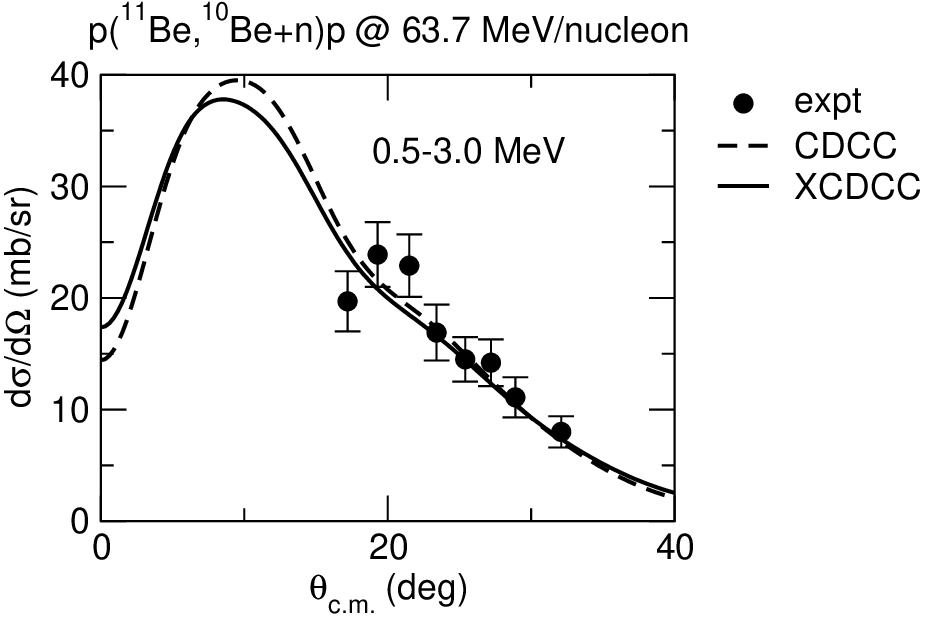}
\caption{\label{FIG:be11-bin1}
$^{11}$Be breakup at 60 MeV/nucleon with the relative energy
between breakup fragments in the range 0.5--3.0 MeV. Experimental data are from Ref.~\cite{shrivastava}.}
\end{figure}
\begin{figure}
\includegraphics[height=5cm]{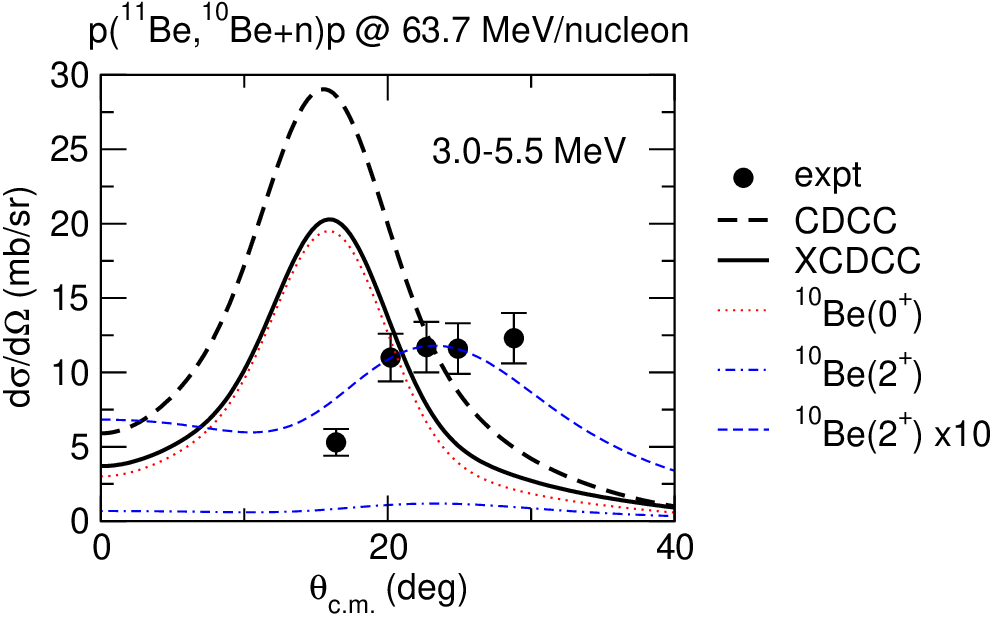}
\caption{\label{FIG:be11-bin2} (Color online)
$^{11}$Be breakup at 60 MeV/nucleon with the relative energy
between breakup fragments in the range 3.0--5.5 MeV. Experimental data are from Ref.~\cite{shrivastava}.}
\end{figure}
The breakup angular distribution data and the associated theory prediction for the lower energy bin (0.5--3.0 MeV) and the higher energy bin (3.0--5.5 MeV) are presented in Figs.~\ref{FIG:be11-bin1} and  \ref{FIG:be11-bin2} (the equivalent of Figures 3b and 3c of Ref.~\cite{shrivastava}).
Firstly, the CDCC calculations of Ref.~\cite{shrivastava} were redone, with a higher CDCC bin density.
We find that converged CDCC calculations, for the lower energy bin, do in fact agree well with the data (dashed line in Fig.~\ref{FIG:be11-bin1}), contrary to what is presented in Ref.~\cite{shrivastava}. Results do not change significantly when core excitation is included with XCDCC (solid line in Fig.~\ref{FIG:be11-bin1}), as could be expected expected due to the resonance structure
in this energy region. One can conclude that a single-particle description for
the first $d_{5/2}$ resonance is adequate.

The data for the higher energy bin is not well described within the
single particle CDCC model (dashed line in Fig.~\ref{FIG:be11-bin2}).
To see if this discrepancy can be explained by excited core contributions, we include the excited $^{10}$Be($2^+$) components in the reaction mechanism, within XCDCC (solid line).
As shown in Fig.~\ref{FIG:be11-bin2}, core excitation lowers the cross section
but does not significantly change the shape of the distribution. It becomes clear that core excitation does not help to reproduce the shape of the higher energy angular distribution.
The main reason for this is that for the $^{11}$Be coupled channel model of \cite{be11nunes},
most of the breakup ends up in the $^{10}$Be ground state. Fig.~\ref{FIG:be11-bin2}
also shows the breakup cross section to the $0^+$ and $2^+$ states of $^{10}$Be
(red/dotted line and the dot-dashed/blue line respectively). We see that whereas the ground state distribution has the original shape of the CDCC calculation, the shape of the distribution to the excited state reproduces the data (to illustrate this fact, we show the breakup cross section to $^{10}$Be$(2^+)$ multiplied by 10 by the dashed/blue line).
The reason for this maybe that the large number of resonances in this region are not reproduced well by our particle-rotor model for $^{11}$Be.
The only resonance that appears in this model is the $3/2^+$, for which the width is not narrow enough to attract significant cross section.
Some suggest that more exotic structures are responsible for resonances in this region \cite{cappuzzello}.
Without exotic resonances built on excited core components in our structure model, the breakup cross section is still dominated by ground state $^{10}$Be fragments.

\section{\label{conclusion} Concluding remarks}

A consistent analysis of reactions involving the halo nucleus $^{11}$Be on protons, at two intermediate energies ($\sim$40 and $\sim$60 MeV/nucleon) are performed and compared with data.
An optical model approach, based on a cluster folding potential constructed from the $^{10}$Be+p potential fitted to the appropriate elastic data, is unable to describe the $^{11}$Be elastic data.
The inclusion of breakup effects improve the description,
but theoretical predictions still overestimate the elastic cross section at larger scattering angles.
The inclusion of core excitation does not affect the elastic distribution significantly.
Note however that these results include no artificial renormalization of the optical potential.
Elastic scattering experiments with radioactive beams at large facilities have repeatedly been undermined. 
The fact that the best reaction models are still unable to fully describe the mechanisms for the $^{11}$Be case, shows the need for a more varied and better elastic scattering experimental program.

In this work we also study the breakup channel explicitly.
Core excitation in the description of the continuum, within XCDCC,
produces a slight modification of the distribution.
These breakup calculations are compared to the data at $63.7$ MeV/nucleon, for two energy bins 0.5--3.0 MeV and 3.0--5.5 MeV.
For the lower energy bin, the shape of the angular distribution is well reproduced by the models.
The same cannot be said for the higher energy bin.

The XCDCC calculations predict breakup states to specific states of the core $^{10}$Be.
This level of detail is still not available in the data, but it could be helpful information, even
at an integrated level, to identify possible causes for the remaining disagreement with 
the data.  

Another important point is related to the basis used to describe the breakup states.
As discussed in Ref.~\cite{moro}, within CDCC,  one can describe the three body final state
continuum $^{10}$Be+n+p in the $^{11}$Be continuum basis or in the deuteron continuum basis.
In this work we used the $^{11}$Be basis. Work in Ref.~\cite{moro} shows that in practice the two choices do not provide the same result. Efforts are underway to tackle this problem within a Faddeev framework \cite{fonseca}.
These results may have important implications to the theory-experiment mismatch.

\section*{Acknowledgements}

We thank the high performance computing center (HPCC) at MSU for
the use of their facilities. This work is supported by NSCL,
Michigan State University, the National Science Foundation through grant PHY-0555893,
by NNSA through US DOE Cooperative Agreement DEFC03-03NA00143 at Rutgers University,
and by the US DOE under contract No. DE-FG02-96ER40983 at the University of Tennessee.

\end{document}